\documentclass[draft,12pt]{iopart}

\input BoxedEPS
\SetRokickiEPSFSpecial  %% for dvips by Tom Rokicki, for VMS

\def\mSe{ce}\def\mSo{se}
\def\mJe{Je}\def\mJo{Jo}
\def\mYo{Yo}

\def\mHo{Ho}
\def\mIo{Io}
\def\mKo{Ko}
\def\ynm{{\cal Y}^m_n}
\def\ynz{{\cal Y}^0_n}

\HideDisplacementBoxes

\begin{document}

\title{Casimir Energies and General Relativity Energy Conditions}
\author{Noah Graham}
\ead{ngraham@middlebury.edu}
\address{Department of Physics, Middlebury College  Middlebury, VT 05753}

\begin{abstract}
Quantum systems often contain negative energy densities.  In general
relativity, negative energies lead to time advancement, rather than
the usual time delay.  As a result, some Casimir systems appear to
violate energy conditions that  would protect against exotic phenomena
such as closed timelike curves and superluminal travel.  However, when
one examines a variety of Casimir systems using self-consistent
approximations in quantum field theory, one finds that a particular
energy condition is still obeyed, which rules out exotic phenomena.  I
will discuss the methods and results of these calculations in detail
and speculate on their potential implications in general relativity.
\end{abstract}


\maketitle

\section{Introduction}

General relativity allows spacetimes of any geometry.  Given any
$g_{\mu\nu}$, we can compute $R_{\mu\nu}$ and $R$, and then 
set up a matter configuration whose stress-energy tensor is
\begin{equation}
T_{\mu\nu} = \frac{1}{8\pi G}\left( R_{\mu\nu} - \frac{1}{2}
g_{\mu\nu}R \right)
\end{equation}
to obtain a solution to Einstein's equations with the desired geometry.
As a result, nothing seems to prohibit the existence of exotic phenomena
such as closed timelike curves \cite{cpc}, traversable wormholes
\cite{Morris88b}, or superluminal travel \cite{Olum:1998mu}.
Therefore we expect there to exist some restrictions the possible tensors
$T_{\mu\nu}$, usually called energy conditions, which would lead to
restrictions on the possible spacetime geometries.  While it is
straightforward to show that classical field theories obey energy
conditions that are strong enough to forbid exotic phenomena,
quantum field theories appear to violate these conditions.

We will focus on the following energy conditions:
\begin{itemize}
\item
The Weak Energy Condition (WEC) requires that for
all timelike vectors
$V^\mu$, 
\begin{equation}
T_{\mu\nu} V^\mu V^\nu \ge 0
\label{WEC}
\end{equation}
That is, all observers see positive energy density.

\item
The Null Energy Condition (NEC) is
weaker than the WEC, and requires that Eq.~(\ref{WEC})
hold only for null vectors $V^\mu$.

\item
The Averaged Null Energy Condition (ANEC) is
weaker than the NEC, and only requires that the NEC
hold when integrated over a complete null geodesic.

\end{itemize}

In a classical background, all of these conditions can be imposed
consistently, and any one of them would be sufficient to rule out
exotic phenomena \cite{cpc}.  On the other hand, the standard Casimir
system of parallel conducting plates has a static negative energy
between the plates, and thus it appears to violate all of these
conditions.  However, there are some obvious caveats to this result.
An external agent has to hold the plates apart, and impose the
boundary condition.  Furthermore, in the case of ANEC, the geodesic has
to go through the boundary, where we are ignoring the effects of the
material.  Since gravity couples to all sources of stress-energy, 
we cannot consider the contribution of the Casimir energy without also
including the contributions due to the materials and their internal
interactions.

\begin{figure}[htbp]
\centerline{\BoxedEPSF{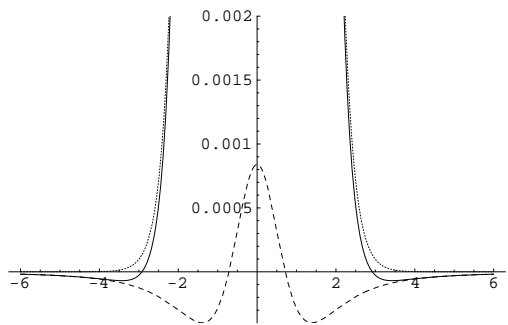 scaled 650}}
\caption{\label{wall}
Classical energy density due to the wall (dotted), the
quantum correction (dashed), and the total (solid), for typical values
of the coupling constants..}
\end{figure}

Previous work \cite{Graham:2002yr,Olum:2002ra} avoided these problems
by considering a domain wall background that is renormalized with
standard counterterms.  In this way, all the contributions to the
stress-energy tensor can be included.  The result, shown in
Figure~\ref{wall}, is that for any coupling, there is always a small
region at large enough $x$ where the energy density is negative.  So
we can always choose the coupling small enough that our semiclassical
approximation is reliable and still find negative energies.  This
situation also violates quantum inequalities \cite{F&Ro95,F&Ro97,Pfen98a}.
Although this system violates WEC, since we have a region of negative
energy density, and NEC, if we take a null vector in the
region of negative energy pointing perpendicular to the wall, ANEC is
still obeyed, since the contribution from the complete
geodesic perpendicular to the wall is only negative if the total
energy of the domain wall is negative (indicating an unstable vacuum
and a breakdown of our approximation).

In a number of other examples, explicit calculation shows
that ANEC is obeyed.  These include a geodesic outside a 
spherically symmetric background potential \cite{Perlov:2003}
or a dielectric sphere \cite{Graham:2004}. Other calculations
also show that energy condition violation is more difficult to achieve
in realistic situations than idealized models would suggest
\cite{Sopova:2002cs,Sopova:2005}.  ANEC is also known to be obeyed by
free scalar \cite{Klinkhammer} and electromagnetic
\cite{Folacci:1992xg} fields in flat spacetime. Other works have found
restrictions on energy condition violation in flat space
\cite{Borde01,F&Ro95,F&Ro96b}.

\section{Plate with a hole calculation}

To further investigate ANEC, work done in collaboration with Ken Olum
\cite{plate} considers another alternative.  We avoid the plate by
drilling a hole in it for the geodesic to travel through.  Then we
might expect that the region around the hole would provide only a
small correction to the negative contribution from the rest of the
geodesic, yielding a violation of ANEC.  Since the geodesic never
encounters the material, the result will be finite with no
contributions from the counterterms (which have support only where
there is a potential).  Our approach will be to use a
Babinet's principle argument to show that the Casimir energy of a
Dirichlet plate with a hole is the sum of the Casimir energy of a full
Dirichlet plate and the Casimir energy of a Neumann disk.  Then we
will use scattering theory in ellipsoidal coordinates to solve the
disk problem, and add in the standard result for the full plate.

\begin{figure}[htbp]
\centerline{\BoxedEPSF{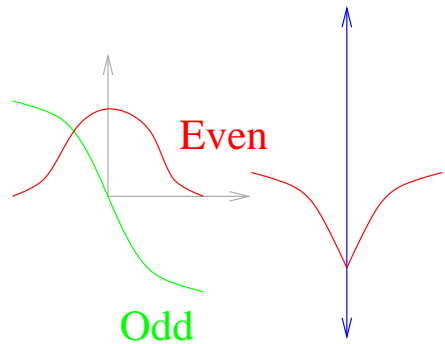 scaled 1000}}
\caption{\label{bab2}
In free space (left) we have even
and odd normal mode wave functions.  In the presence of a Dirichlet
plate (right), the even functions are replaced by the odd functions with a
change of sign crossing the plate.}
\end{figure}

In free space, we can decompose the spectrum into modes that are odd
or even under reflection of the $z$-axis.  Imposing a Dirichlet
boundary condition at $z=0$ has no effect on the odd modes, since they
already obey the boundary condition.  However, the even modes are
modified:  They turn into the odd modes multiplied by the sign of $z$,
with a cusp at $z=0$, as shown in Figure~\ref{bab2}.

\begin{figure}[htbp]
\centerline{\BoxedEPSF{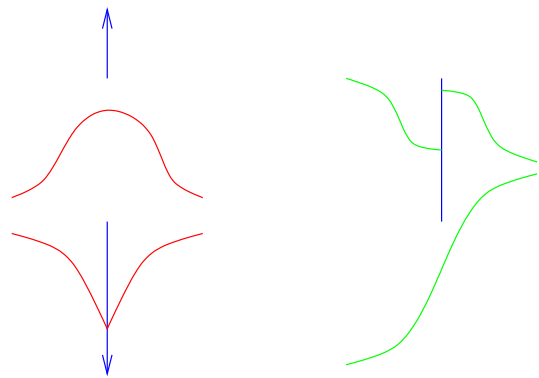 scaled 1000}}
\caption{\label{bab1}
If there is a hole in the Dirichlet plate (left), the new even
functions satisfy Neumann conditions in the hole and Dirichlet
conditions elsewhere.  The odd functions for a Neumann patch with the
same shape as the hole (right) are the same except for a change of
sign between sides.}
\end{figure}

If there is a hole in the plate, we find even functions that obey Neumann
conditions in the hole (since they are even) and Dirichlet conditions
elsewhere.  Up to a similar sign flip between positive and negative
$z$, these are the same functions as one obtains for the odd modes of a
Neumann patch with the same shape as the hole, as shown in Figure~\ref{bab1}.

Let $E$ be the contribution to the Casimir energy from the 
free even modes, $O$ be the contribution to the Casimir energy 
from the free odd modes, and $A$ be the contribution to the Casimir
energy from the modes for the plate with the hole.  Then we have the
following total contributions to the energy, where we have taken the
difference with the free space result:

\begin{table}[htbp]
\begin{tabular}{lccc}
\hbox{Free space:} &
$E + O$
&${\hbox{\small subtract free space} \atop \Longrightarrow}$
& zero \cr
\hbox{Dirichlet plate:} &
$O + O$
&${\hbox{\small subtract free space} \atop \Longrightarrow}$&
$O - E$ \cr
Dirichlet plate with hole: &
$A+O$ \quad
&${\hbox{\small subtract free space} \atop \Longrightarrow}$&
$A-E$ \cr
Neumann disk: &
$E + A$
&${\hbox{\small subtract free space} \atop \Longrightarrow}$&
$A-O$
\end{tabular}
\end{table}

Thus we have found
\begin{equation}
\hbox{[Dirichlet plate with a hole] =
[Neumann disk] + [Dirichlet plate]} \,.
\end{equation}

Next, we set up the disk problem.  We will begin in two space dimensions
and find the Casimir energy of a Neumann line segment.  For notational
consistency with the three-dimensional case, we consider the $x$-$z$
plane with the geodesic running along the $z$-axis and the segment
running along the $x$-axis from $x=-d$ to $x=d$.  We define
elliptical coordinates $\mu$ and $\theta$ by
\begin{equation}
x = d\cosh\mu\cos\theta \qquad
z = d\sinh\mu\sin\theta
\end{equation}
and then the boundary condition is applied at the points where the radial
coordinate $\mu$ is zero.  In elliptical coordinates, the problem is
still separable, but there is no analog of angular momentum
conservation.  We still have separate angular and radial functions,
but we now have the additional dimensionful parameter $d$.  Thus the
angular functions are now not simply functions of $\theta$; they can
now depend also on $kd$, where $k$ is the wave number.  Similarly the
radial functions can depend on $kd$ and $r/d$ separately, rather than
depending simply on the product $kr$.

As described above, we have normal mode solutions that are even and
odd under reflection of the $z$-axis.  Since the line segment has
Neumann boundary conditions, only the odd modes need to be modified
from the free case.  These become
\begin{equation}
\psi o_m(\mu, q)=\frac{1}{2}
\left[e^{2i\delta}\mHo^{(1)}_m (\mu, q)+\mHo^{(2)}_m (\mu, q)\right]
\mSo_m(\theta, q)
\end{equation}
where $q=(dk/2)^2$.  Here $\mSo_m(\theta, q)$ is the odd angular
Mathieu function and $\mHo^{(1)}_m$ and  $\mHo^{(2)}_m$ are the
corresponding odd radial functions of the third and fourth kinds, with
$\mHo^{(1)}_m = \mJo_m + i \mYo_m$  and $\mHo^{(2)}_m = \mJo_m  - i
\mYo_m$, where $\mJo_m$ and $\mYo_m$ are the radial functions of the
first and second kind respectively.  The phase shift $\delta(q)$ is
defined by imposing the boundary condition, which yields
\begin{equation}
e^{2i\delta}= -\frac{\mHo^{(2)}_m {}' (0, q)}{\mHo^{(1)}_m {}' (0, q)}
\end{equation}
where the derivative is with respect to $\mu$.  

We have adopted the normalization conventions of \cite{Abramowitz}, but
have modified their notation to make the analogy to the circular case
clearer.  In these conventions the even and odd angular functions
$\mSe_m(\theta,q)$ and $\mSo_m(\theta,q)$ are normalized just like the
$\cos m\theta$ and $\sin m\theta$ solutions in the circular 
case, so that 
\begin{equation}
\int_0^{2\pi} d\theta \mSe_m(\theta, q)^2=\int_0^{2\pi} d\theta
\mSo_m(\theta, q)^2 =\pi \,.
\end{equation}
As in the circular case, 
$m=0,1,2,3\ldots$ for the even functions\footnote{In the circular
case, for $m=0$ the even solution is a constant, which is normalized
to be $\frac{1}{\sqrt{2}}$ rather than $\cos 0 = 1$, so that its
normalization is consistent with the other modes.}
and $m=1,2,3\ldots$ for the odd functions.  The radial
functions are then normalized so that they approach the corresponding Bessel
functions at large radius.\footnote{Note that the functions $Je_m$ and
$Jo_m$ defined in \cite{MorseFeshbach} have an additional factor of
$\sqrt{\pi/2}$.}  Instead of singularities at zero radius as in the
circular case, the radial functions of the second kind have jump
discontinuities --- the singularity is now ``spread'' over the
interfocal separation.

From these wavefunctions, we form the normalized quantum field $\phi$
\begin{eqnarray}
\hspace*{-2cm}
\phi(\mu,\theta)=
\sum_{m=0}^\infty \int_0^\infty dk\,\sqrt{\frac{k}{2\pi\omega}}
\left( \mJe_m(\mu, q) \mSe_m (\theta, q) \hat b_{k}^{m}{}^\dagger
+\mJo_m(\mu, q) \mSo_m (\theta, q) \hat c_{k}^{m}{}^\dagger\right)
e^{i\omega t} \cr
+ \hbox{~complex conjugate}
\end{eqnarray}
where the odd term vanishes for $m=0$.  The stress-energy tensor for a
minimally-coupled scalar field is
\begin{equation}
\label{eqn:stressenergytensor}
T_{\lambda\nu}=\partial_\lambda\phi\partial_\nu\phi - 
\frac{1}{2}\eta_{\lambda\nu}
\left[\partial^\lambda\phi\partial_\lambda\phi\right]\,.
\end{equation}
For a null vector,
$\eta_{\lambda\nu}V^\lambda V^\nu= 0$, so we have
\begin{equation}
T_{\lambda\nu}V^\lambda V^\nu= \left(V^\alpha\partial_\alpha\phi \right)^2\,.
\end{equation}
Taking our geodesic along the $z$-axis, $V = (1,{\bf \hat z})$, we have
\begin{equation}
\label{eqn:perpendicularANEC}
T_{\lambda\nu}V^\lambda V^\nu =  \dot\phi^2+ \left(\partial_z\phi\right)^2\,.
\end{equation}
We then substitute the result for the quantum field into this expression,
subtract the free space result, and take the vacuum expectation value
\cite{Bordag:1996,Saharian:2000mw,Graham:2002cas,Perlov:2003}.
For computational efficiency, we extend the $k$ integration to the
whole real axis and use contour integration to obtain an integral over
the imaginary $k$-axis, using $k=i\kappa$.  We obtain \cite{plate}
\begin{equation}
\langle\dot\phi^2\rangle 
= \frac{1}{\pi^2}\sum_{m=1}^\infty\int_0^\infty d\kappa
\frac{\mIo_m'(0,\varphi)}{\mKo_m'(0,\varphi)}
\kappa^2 \mKo_m(\mu,\varphi)^2\mSo_m(\theta,-\varphi)^2 \,.
\end{equation}
On the $z$-axis, terms with $m$ even vanish, so we have
\begin{equation}
\langle\dot\phi^2\rangle =
\frac{1}{\pi^2}\sum_{m=1}^\infty{}' \int_0^\infty d\kappa
\frac{\mIo_m'(0,\varphi)}{\mKo_m'(0,\varphi)}
\kappa^2 \mKo_m(\mu,\varphi)^2\mSo_m(\pi/2,-\varphi)^2
\end{equation}
where the prime on the summation sign indicates that we sum over odd
values of $m$.  The sum over channels and integral over $k$ are
absolutely convergent, since we have subtracted the contribution of the free
theory and we are away from the interactions, where all the remaining
counterterms vanish.  Similarly, on the $z$-axis we have
\begin{equation}
\hspace{-1.25cm}
\langle(\partial_z\phi)^2\rangle
=-\frac{1}{\pi^2 d^2 \cosh^2 \mu}\sum_{m=1}^\infty {}' \int_0^\infty d\kappa
\frac{\mIo_m'(0,\varphi)}{\mKo_m'(0,\varphi)}
\mKo_m'(\mu,\varphi)^2\mSo_m(\pi/2,-\varphi)^2\,.
\end{equation}

In three dimensions, we employ oblate ellipsoidal coordinates,
\begin{equation}
\hspace{-1cm}
x = d\sqrt{(\xi^2+1) (1-\eta^2)}\cos\phi \quad
y = d\sqrt{(\xi^2+1) (1-\eta^2)}\sin\phi \quad
z = d\eta\xi
\end{equation}
using the conventions of \cite{Meixner}.  As in spherical coordinates, the
solutions are indexed by $n=0,1,2,3\ldots$ and $m=-n \ldots n$.  The
angular solutions are spheroidal harmonics $\ynm(ic; \eta,
\phi)$, where $c=kd$, $\cos^{-1} \eta$ is the polar angle, and $\phi$
is the azimuthal angle.  These are normalized analogously to ordinary
spherical harmonics. The radial functions are $R^{m(1)}_n(ic;-i\xi)$
and $R^{m(2)}_n(ic;-i\xi)$, where $\xi$ is the radial coordinate.
These are normalized so that they approach the usual spherical Bessel
functions at large radius.  The factors of $\pm i$ convert from prolate to
oblate coordinates.

Now we consider Neumann boundary conditions on the disk
$\xi = 0$.  If $m+n$ is even, the functions obey the boundary conditions
already.  Otherwise we need to take a combination of $R^{m(1)}_n$ and
$R^{m(2)}_n$.  With $R^{m(3)}_n=R^{m(1)}_n+iR^{m(2)}_n$ and
$R^{m(4)}_n=R^{m(1)}_n-iR^{m(4)}_n$ we can write the desired radial
function
\begin{equation}
\psi^m_n(ic;-i\xi)=\frac{1}{2}
\left[e^{2i\delta(ic)}R^{m(3)}_n(ic;-i\xi) + R^{m(4)}_n(ic;-i\xi)\right]
\end{equation}
with the condition
\begin{equation}
e^{2i\delta(ic)}= -\frac{R^{m(4)}_n{}'(ic;0)}{R^{m(3)}_n{}'(ic;0)}
\end{equation}
where the derivative is with respect to the second argument.

Passing to the imaginary $k$-axis in the same way as in the two-dimensional
case, we obtain \cite{plate}
\begin{equation}
\hspace*{-1cm}
\langle\dot\phi^2\rangle = -\frac{1}{\pi}
\sum_{n=0}^\infty\sum^n_{m=-n}{}' \int_{0}^\infty d\kappa\,
\kappa^3\frac{R^{m(1)}_n {}'(\gamma;0)}{R^{m(3)}_n {}'(\gamma;0)}
|\ynm(\gamma;\eta,\phi)|^2 R^{m(3)}_n(\gamma;-i\xi)^2
\end{equation}
where $\gamma = ic = ikd = -\kappa d$
and the prime on the summation sign means that only odd values of
$m+n$ are included.  On the axis, we have
\begin{equation}
\langle\dot\phi^2\rangle =
-\frac{1}{\pi} \sum_{n=1}^\infty{}' \int_{0}^\infty d\kappa\,
\kappa^3\frac{R^{0(1)}_n {}'(\gamma;0)}{R^{0(3)}_n {}'(\gamma;0)}
|\ynz(\gamma;1,\phi)|^2 R^{0(3)}_n(\gamma;-i\xi)^2
\end{equation}
where we have specialized to $m = 0$ because the contributions from
nonzero $m$ vanish on the axis, leaving only a sum over odd values
of $n$.  Similarly, on the axis we have
\begin{equation}
\langle (\partial_z \phi)^2\rangle = -\frac{1}{\pi d^2}
\sum_{n=1}^\infty{}' \int_{0}^\infty  d\kappa \,
\kappa \frac{R^{0(1)}_n {}'(\gamma;0)}{R^{0(3)}_n {}'(\gamma;0)}
|\ynz(\gamma;1,\phi)|^2 R^{0(3)}_n{}'(\gamma;-i\xi)^2
\end{equation}
where the primes on the radial functions indicate derivatives with
respect to the second argument.

Once we have computed the result for a Neumann segment and disk, we must
simply combine with the Dirichlet mirror result (see for example
\cite{Mostepanenko97}),
\begin{equation}
\langle\dot\phi^2\rangle + \langle(\partial_z \phi)^2\rangle =
\left\{ \begin{array}{l@{\quad\quad}l}
\displaystyle
- \frac{1}{32 \pi z^3} & \hbox{in two dimensions, and} \cr
\displaystyle
- \frac{1}{16 \pi^2 z^4} & \hbox{in three dimensions}
\end{array}\right. \,.
\end{equation}

\section{Results}
\begin{figure}
\BoxedEPSF{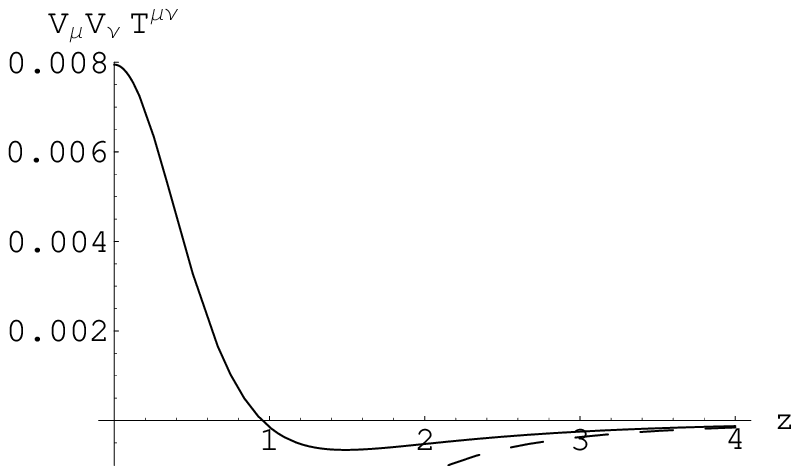 scaled 900}\hfill
\BoxedEPSF{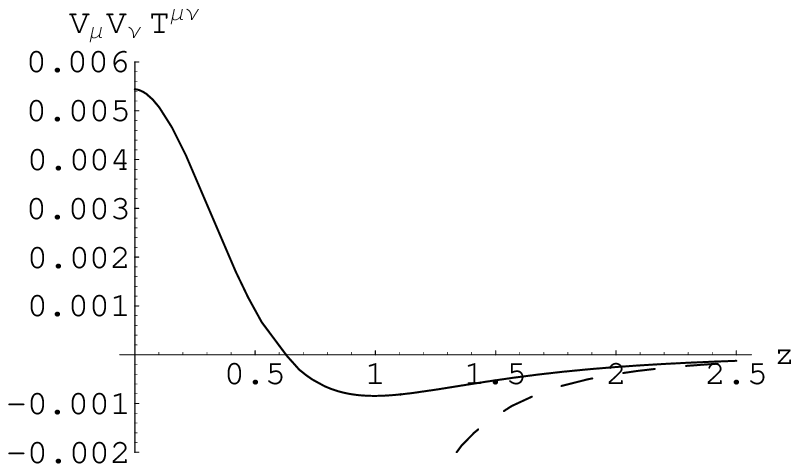 scaled 900}
\caption{\label{nec}
Contributions to NEC in two dimensions (left) and three dimensions
(right) for a Dirichlet plate with a hole of unit radius, as functions
of distance along the axis passing through the center of the hole.
Extrapolation is used for points at a distance less than $0.15$ 
in the left panel and $0.25$ in the right panel.
The dotted lines show the perfect mirror result.}
\end{figure}

We carried out the two-dimensional calculations using the C++ Mathieu
function package of Alhargan \cite{Alhargan:2000,Alhargan:2000a}, with
minor enhancements to accommodate the extreme range of Mathieu
functions needed to accurately compute this sum.  We have also adapted
the code to use our normalization conventions instead of those of
\cite{MorseFeshbach}.  The sums and integrals are then done by calling
the C++ code from Mathematica routines.\footnote{Mathematica does
provide built-in support for both radial and angular Mathieu
functions, but only for functions of the first kind (as of version
5.2).}  In three dimensions, we used the Mathematica spheroidal packages
of Falloon \cite{Falloon:2002}, with minor enhancements to avoid
memory leaks, allow compatibility with current versions of
Mathematica, and improve efficiency for our application.  Results in
two and three dimensions are shown in Figure~\ref{nec}.  We see that
in both cases the hole has a dramatic effect, overwhelming the
NEC-violating behavior away from the plate so that ANEC is obeyed.
Although all field theory divergences are well under control, this
calculation is still highly nontrivial for points near the hole,
because both the perfect mirror and the disk have energies that
diverge like $1/z^{n+1}$ where $n$ is the space dimension and $z$ is
the distance to the hole.  The final result, however, is perfectly
finite  --- the origin is just a point in empty space. 
As a result, we have to stop our calculation a fixed distance
away from the origin, depending on our numerical
precision (and patience).  We then extrapolate the result to zero, and
verify that it goes to a finite result with zero slope (without
building this requirement into the extrapolation).  Far away, the
calculation approaches the perfect mirror result, which is also shown
in Figure~\ref{nec}.

\begin{figure}
\BoxedEPSF{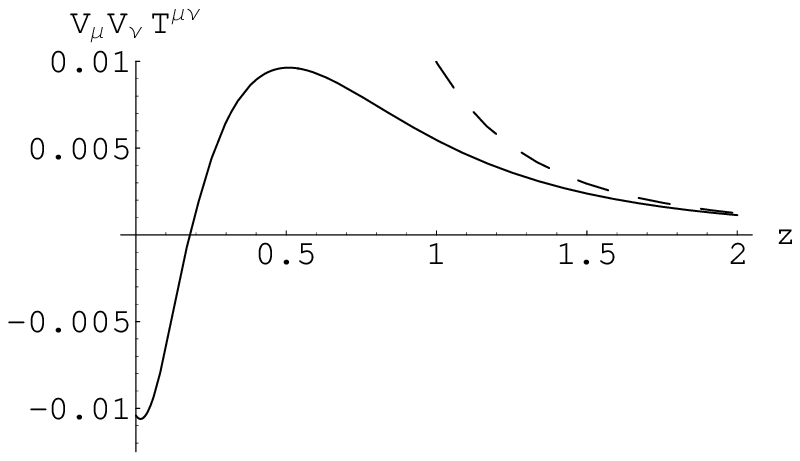 scaled 900}\hfill
\BoxedEPSF{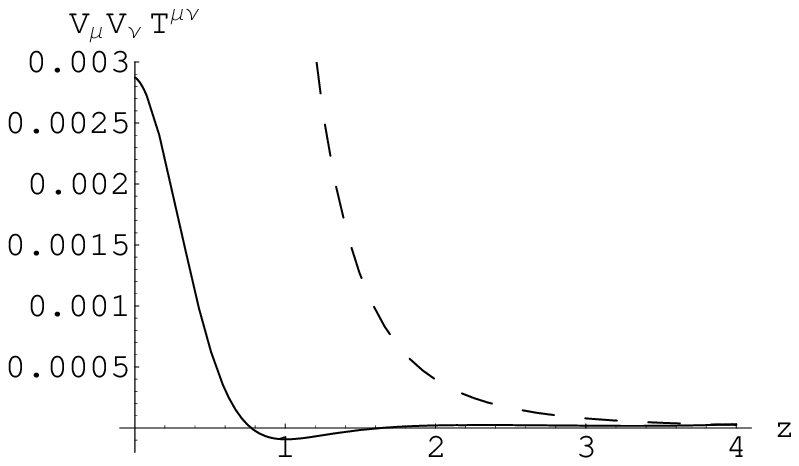 scaled 900}
\caption{\label{necn}
Contributions to NEC in two dimensions (left) and three dimensions
(right) for a Neumann plate with a hole of unit radius, as functions
of distance along the axis passing through the center of the hole.
Extrapolation is used for points at a distance less than $0.11$ 
in the left panel and $0.25$ in the right panel.
The dotted lines show the perfect mirror result.}
\end{figure}

Since the Dirichlet plate with a hole obeys ANEC, and the Neumann and
Dirichlet cases typically contribute with opposite signs, we might
then expect a Neumann plate to violate ANEC, with a negative
contribution from the hole overwhelming the positive contribution from
far away.  It is straightforward to repeat this analysis for that case
\cite{plate}. However, as shown in Figure~\ref{necn}, we see that ANEC
is also obeyed for a Neumann plate with a hole.  Furthermore, both the
Dirichlet and Neumann results extend to the case of two plates in the
limit of both large and small holes \cite{plate}.

\section{Conclusions}
Energy condition violation by Casimir systems initially looks
dramatic, but in consistent field theory models it is
modest or nonexistent.  We do not have an example of ANEC violation in
flat space --- that is, due to quantum effects in a background
of non-gravitational quantum fields.  Violations are known to exist
\cite{Viss96a,Viss96b,Viss96c,Viss97a} in curved space, when one
considers the quantum effects of curvature caused by distant masses.
But the magnitude of this violation is typically much smaller
than competing effects associated with the source of the curvature
itself (unless we consider Planck-scale objects, where classical
general relativity is unreliable), so it is not clear that this
violation will persist in the full theory.  If ANEC (possibly with
appropriate modifications in curved space) is always obeyed by
realistic quantum field theories with no uncontrolled external agents,
it would prevent the appearance of exotic phenomena.  Work is
underway to investigate the possibility of finding analytic arguments
to ensure that ANEC is always obeyed in flat space.

\section*{Acknowledgements}
This work was done in collaboration with Ken Olum (Tufts).
N.~G. was supported in part by a Cottrell College Science Award from
Research Corporation and by a Baccalaureate College Development grant
from the Vermont Experimental Program to Stimulate Competitive
Research (VT-EPSCoR).

\section*{References}

\bibliographystyle{unsrt}
\bibliography{gr}

\end{document}